\documentclass[
 reprint,
 amsmath,amssymb,
]{revtex4-1}

\usepackage{graphicx}
\usepackage{dcolumn}
\usepackage{bm}

\setcitestyle{super}

\usepackage[
text={7.4in,10in},centering,
]{geometry}

\begin{document}

\title{\Large\sffamily Attosecond streaking of photoelectron emission from disordered solids}

\author{W. A. Okell$^{1}$}
\email{william.okell09@imperial.ac.uk}
\author{T. Witting$^{1}$}
\author{D. Fabris$^{1}$}
\author{C. A. Arrell$^{1,2}$}
\author{J. Hengster$^{3}$}
\author{S. Ibrahimkutty$^{4}$}
\author{A. Seiler$^{4}$}
\author{M. Barthelmess$^{5}$}
\author{\mbox{S. Stankov$^{4}$}}
\author{D. Y. Lei$^{1,6}$}
\author{Y. Sonnefraud$^{1,7}$}
\author{M. Rahmani$^{1}$}
\author{Th. Uphues$^{3}$}
\author{S. A. Maier$^{1}$}
\author{J. P. Marangos$^{1}$}
\author{J. W. G. Tisch$^{1}$}
\affiliation{
$^{1}$Blackett Laboratory, Imperial College London, Prince Consort Road, London SW7 2AZ, UK}
\affiliation{
$^{2}$Current address: Laboratory of Ultrafast Spectroscopy, ISIC, Ecole Polytechnique F\'ed\'erale de Lausanne, CH-1015 Lausanne, Switzerland
}
\affiliation{
$^{3}$Center for Free-Electron Laser Science, Attosecond Research and Science Group, Hamburg University, Luruper Chaussee 149, 22761 Hamburg, Germany}
\affiliation{
$^{4}$IPS, Karlsruhe Institute of Technology, Hermann von Helmholtz Platz 1, 76344 Eggenstein-Leopoldshafen, Germany
}%
\affiliation{
$^{5}$Center for Free-Electron Laser Science, DESY, Notkestraße 85, 22607 Hamburg, Germany}
\affiliation{
$^{6}$Current address: Department of Applied Physics, The Hong Kong Polytechnic University, Hong Kong, China}
\affiliation{
$^{7}$Current address: CNRS, Institut N\'EEL,  F-38042 Grenoble, France}

\date{\today}
\maketitle

\bf Attosecond streaking of photoelectrons emitted by extreme ultraviolet light has begun to reveal how electrons behave during their transport within simple crystalline solids. Many sample types within nanoplasmonics, thin-film physics, and semiconductor physics, however, do not have a simple single crystal structure. The electron dynamics which underpin the optical response of plasmonic nanostructures and wide-bandgap semiconductors happen on an attosecond timescale. Measuring these dynamics using attosecond streaking will enable such systems to be specially tailored for applications in areas such as ultrafast opto-electronics. We show that streaking can be extended to this very general type of sample by presenting streaking measurements on an amorphous film of the wide-bandgap semiconductor tungsten trioxide, and on polycrystalline gold, a material that forms the basis of many nanoplasmonic devices. Our measurements reveal the near-field temporal structure at the sample surface, and photoelectron wavepacket temporal broadening consistent with a spread of electron transport times to the surface.\rm

There has been substantial progress in widening the scope of attosecond science since the first attosecond light pulses were measured in 2001\cite{paul_observation_2001,hentschel_attosecond_2001}. Using attosecond streaking, the motion of electrons in atoms can now be resolved with attosecond precision \cite{drescher_time_2002,uphues_ion_2008,uiberacker_attosecond_2007,schultze_delay_2010}. A considerable challenge, however, lies in the application of attosecond streaking to condensed matter systems. Previous attosecond streaking experiments in solids have been carried out only on surfaces with extremely high degrees of atomic structural order and purity, prepared using ion-bombardment, annealing, and in-situ deposition \cite{cavalieri_attosecond_2007,neppl_attosecond_2012}. The challenges associated with sample preparation have limited the progress in attosecond surface science, with only a few studies existing to date \cite{cavalieri_attosecond_2007,neppl_attosecond_2012,magerl_a_2011}. This has furthermore raised the open question whether attosecond streaking could be applied to solid systems with complex composition and morphology, which are often not amenable to surface preparation techniques.  Such samples have generally been considered unsuitable for attosecond streaking because they will typically have disordered (amorphous or polycrystalline) atomic structures and contaminated surfaces.

The interaction between light and metal nanostructures can induce a collective oscillation of the delocalised electrons in the metal. These excitations, known as plasmons, have a diverse range of applications in technologies ranging from metamaterials \cite{shalaev_optical_2007} to bio-sensing \cite{anker_biosensing_2008}. Plasmonics has recently attracted significant interest from the ultrafast physics community \cite{stockman_attosecond_2007,kruger_attosecond_2011,zherebtsov_controlled_2011,herink_field_2012}, in part because the electronic response to the applied light wave can concentrate the electric field into a nanoscopic volume. This can enhance the intensity by tens to hundreds of times, allowing low energy (hundreds of pJ to nJ) laser pulses to access the regime of strong field physics \cite{kruger_attosecond_2011}, where the electrostatic potential of the laser becomes comparable to the electronic binding potential in matter. Furthermore, attosecond streaking provides the unprecedented opportunity to reveal the temporal structure of near-fields in plasmonic nanostructures \cite{stockman_attosecond_2007,skopalova_numerical_2011,sussmann_attosecond_2011}, and could therefore play a crucial role in tailoring the plasmonic response of a system for a specific application. Such techniques are also likely to be useful in the investigation of ultrafast optically induced currents in nanoscopic systems when designing opto-electronic circuits. The ability to perform streaking on disordered samples without prior surface preparation is a prerequisite for streaking measurements on most types of plasmonic sample, since delicate nanostructures will generally be restructured or destroyed by ultrahigh vacuum (UHV) surface preparation techniques, which typically cause morphological changes on nanometre length scales.

Another important area of research is electron transport in solid state physics, particularly in semiconductors. Within the electronics industry, amorphous thin-films are widely used in semiconductor devices such as thin-film transistors \cite{nomura_room_2004}. At the interface with nanoplasmonics, cheap plasmonically enhanced solar cells using amorphous thin-films appear feasible in the near future \cite{atwater_plasmonics_2010}. In semiconductors, electron-hole pair formation, charge density distributions, and their influence on electron propagation, remain largely unexplored experimentally in the attosecond time domain. Measuring the time evolution of these attosecond electron dynamics during the interaction between ultrashort laser pulses and wide-bandgap semiconductors will aid the design of ultrafast opto-electronic circuits \cite{schultze_controlling_2013,krausz_attosecond_2014}. 

In this paper, we present attosecond streaking measurements on structurally disordered thin-films without any \it in-situ \rm UHV surface preparation. We show that robust attosecond-resolved measurements can be performed using streaking on this very general type of sample by fully reconstructing the streaking field at the sample surface with attosecond precision, and by confirming the photoemission of attosecond photoelectron wavepackets. The durations of these wavepackets are consistent with the spread of electron propagation times to the surface associated with a range of emission depths. The materials on which we performed measurements were polycrystalline Au, which is widely used in plasmonics, and amorphous WO$_{3}$ which is a wide-bandgap semiconductor ($3.41\,$eV bandgap \cite{nakamura_fundamental_1981}). Thus, our measurements extend attosecond streaking towards measuring the attosecond time evolution of near-fields in complex opto-electronic systems, and elucidating the electronic dynamics mediating optical coupling to these systems. Our results also indicate that the preparation of samples for attosecond surface science can in many cases be greatly simplified, making the technique more accessible.

In a typical attosecond streaking experiment, a photoelectron wavepacket is emitted by an extreme ultraviolet (XUV) pulse in the presence of a strong near-infrared (NIR) laser field, and the electrons are subsequently accelerated in the field. This process can imprint timing information onto the photoelectron wavepacket due to the well-defined relationship \cite{itatani_attosecond_2002} \mbox{$\boldsymbol{v}_{f}=\boldsymbol{v}_{0}+(e/m_{e})\boldsymbol{A}(t_{i})$} between the final photoelectron velocity $\boldsymbol{v}_{f}$ and the instantaneous vector potential $\boldsymbol{A}(t_{i})$ at the time of photoemission, where $\boldsymbol{v}_{0}$ is the initial photoelectron velocity. The ability to make sub-cycle (with respect to the NIR streaking field) measurements using the technique requires the photoelectron wavepacket to have sub-cycle duration.

A simple estimate of the duration of the photoelectron wavepacket can be made by considering the photoelectron mean free path, and the wavepacket dispersion. For incident photons with energies in the XUV, the maximum depth into the solid from which electrons can be photoemitted is limited by the photoelectron mean free path (typically $<1\,$nm), rather than the photon penetration depth into the solid (typically on the order of $100\,$nm). The valence band photoemission from a solid can be either surface or bulk in origin, with the dominant mechanism being dependent on the bandstructure of the solid \cite{borisov_resonant_2013}. Electrons emitted from the bulk travel, on average, one mean free path to the surface. Assuming perfect screening of the streaking field at the surface, the photoelectrons are streaked with a time delay associated with the transport time to the surface \cite{neppl_attosecond_2012,borisov_resonant_2013}.

For bulk photoelectrons there is, in fact, a range of emission depths, with the main contribution to the total photoelectron yield coming from within one mean free path of the surface. Thus, there is a spread of emission times, and the photoelectron wavepacket emerging from the bulk will be longer than the incident excitation pulse \cite{borisov_resonant_2013}. The temporal broadening of the photoelectron wavepacket compared to the incident XUV pulse is $\sim\!\lambda\sqrt{m_{e}/2\mathcal{E}}$, where $\lambda$ is the mean free path and $\mathcal{E}$ is the photoelectron energy. For a mean free path comparable to the XUV photon penetration depth, the streaking trace would become heavily smeared in the time direction \cite{liao_initial_2014}. For photoelectrons emitted by XUV radiation, the mean free path in solids is generally sub-nanometre \cite{tanuma_experimental_2005}, which will be true for disordered and contaminated samples as well as clean single crystals. For $84\,$eV photoelectrons the mean free paths in Au and WO$_{3}$ are $\sim\!3.9\,\mathrm{\AA}$ and $\sim\!4.9\,\mathrm{\AA}$, respectively \cite{tanuma_experimental_2005}. Over these mean free paths the expected temporal broadening ($72\,$as for Au and $90\,$as for WO$_{3}$) is non-negligible, but sufficiently small for an excitation pulse lasting several hundred attoseconds to emit a sub-cycle photoelectron wavepacket.

The dispersion of the final state will act to further broaden the photoelectron wavepacket in time. The bandstructure of a disordered solid is more complex than for a single crystal, and one might therefore expect a more complex dispersion relation. However, for high energy photoelectrons ($\mathcal{E}\gg W_{f}$, where $\mathcal{E}$ is the photoelectron energy and $W_{f}$ is the work function of the solid) one can, to a first approximation, treat the final electron state as unbound. The free electron group-velocity dispersion is given by
\begin{equation}
\hbar\frac{\partial}{\partial\mathcal{E}}\frac{1}{v_{g}}=-\hbar\sqrt{\frac{m_{e}}{8\mathcal{E}^{3}}},
\end{equation}
where $v_{g}=\sqrt{2\mathcal{E}/m_{e}}$ is the free electron group velocity. The dispersion at $84\,$eV (the typical photoelectron energy in our experiments) is $-72\,\mathrm{as}^{2}\,\mathrm{\AA}^{-1}$, which is small over a sub-nanometre mean free path for an excitation pulse lasting several hundred attoseconds. These simple considerations indicate that even for disordered samples without any surface preparation, it should be possible to photoemit a sub-cycle photoelectron wavepacket using an extreme ultraviolet isolated attosecond pulse lasting several hundred attoseconds. While only clean single crystals have been studied so far, attosecond streaking should be applicable to solids in general, and we next present measurements that confirm this is indeed the case.

\subsection*{\normalsize Results}

\begin{figure}[h!]
\centering
\includegraphics[scale=0.45]{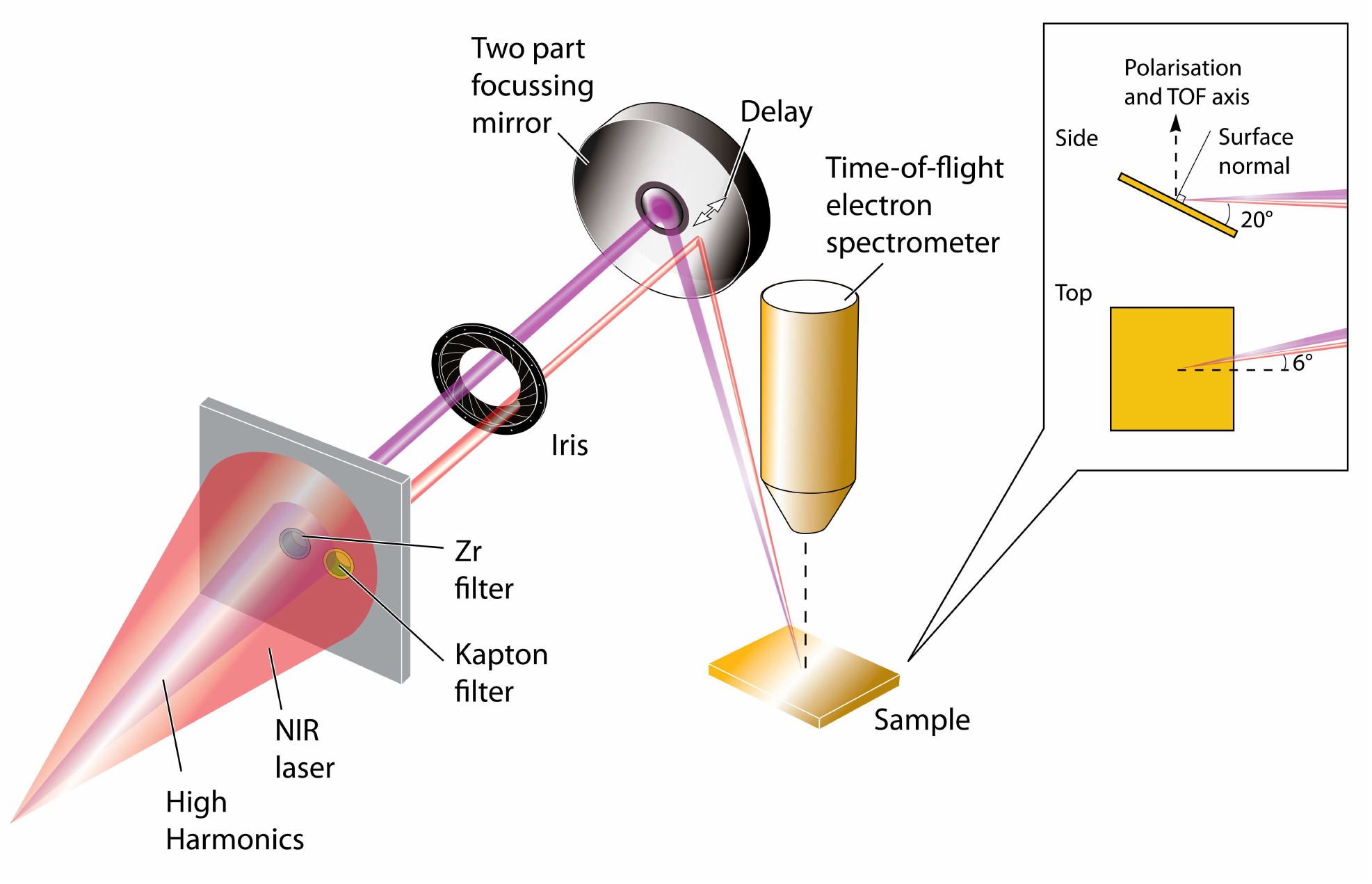}
\caption{\label{setup_fig} Surface streaking experimental setup. NIR and XUV pulses are focused onto the sample with a variable time-delay. Photoemitted electrons are detected with a time-of-flight (TOF) spectrometer. The inset shows the geometry of the incident beam with respect to the sample. The pulses are focused onto the sample with an incidence angle of $\sim\!20\,^{\circ}$. The laser polarisation lies approximately along the TOF axis. The incident beam is also rotated in a horizontal plane by $\sim\!6\,^{\circ}$.}
\end{figure}

\begin{figure}[htb!]
\centering
\includegraphics[scale=0.55]{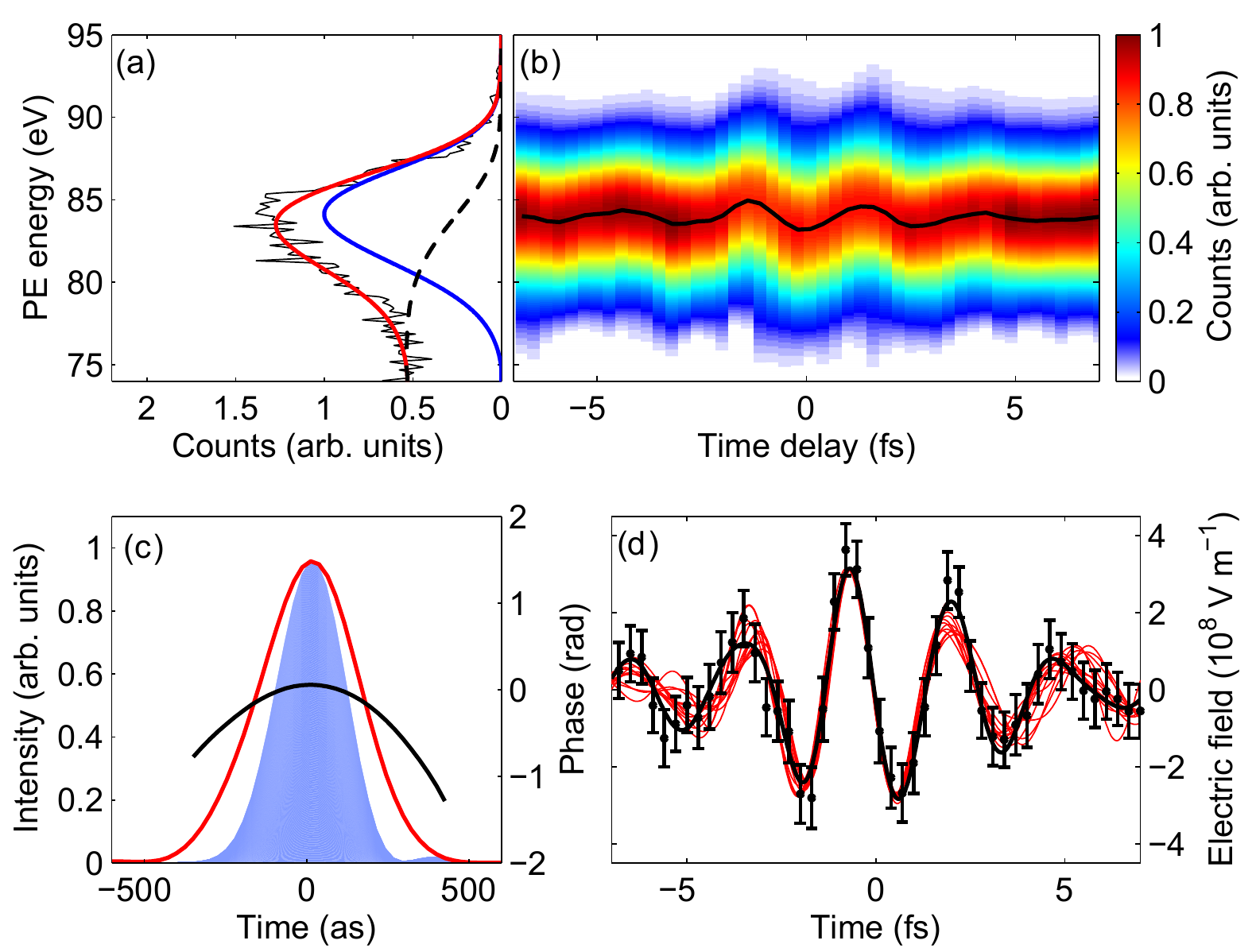}
\caption{\label{W_spec_fig} (a) Unstreaked valence band photoelectron (PE) spectrum of WO$_{3}$, showing raw data (solid black), Fourier filtered spectrum (red), secondary electron background (dashed black) and background subtracted and filtered spectrum (blue). (b) Fourier filtered and background subtracted streaking trace from WO$_{3}$. The delay-dependent central energy of the valence band is shown by the black curve (see methods section). (c) Retrieved photoelectron wavepacket intensity (red) and phase (black) from FROG-CRAB PCGPA algorithm. The shaded area shows the incident XUV pulse. The retrieved wavepacket has a duration of $359_{-25}^{+42}\,$as. (d) Band-pass filtered electric field retrieved from WO$_{3}$ streaking trace (black curve), and unfiltered data points. Retrieved fields from eight separate gas phase streaking measurements \cite{witting_sub_2012} are also shown (red curves). The peak electric field from each gas phase streak has been scaled to the peak field from WO$_{3}$ to aid comparison.}
\end{figure}

To experimentally verify the applicability of attosecond streaking on disordered solids, we performed streaking measurements on two types of sample: an amorphous $20\,$nm-thick film of the semiconductor WO$_{3}$ on a silicon (100) substrate, and a polycrystalline $52\,$nm-thick gold film on silicon (100). Co-propagating  isolated $248\pm15\,$as XUV pulses, and few-cycle sub-$4\,$fs NIR pulses were generated \cite{frank_technology_2012,okell_carrier_2013}, and used to perform attosecond streaking measurements on the metallic films. The experimental setup is shown in figure \ref{setup_fig}. The valence band region of the experimental WO$_{3}$ photoelectron spectrum is shown in figure \ref{W_spec_fig} (a), and displays a clear valence band peak at $84\,$eV. Experimental photoelectron spectra were Fourier filtered, and the secondary electron background from photoelectrons inelastically scattered on their way to the surface was subtracted using a standard approach \cite{shirley_high_1972,neppl_attosecond_2012}.

The filtered and background-subtracted attosecond streaking trace from WO$_{3}$, acquired using $300\,$as time-delay steps and $120\,$s integration time ($1.2\times10^{5}$ laser shots) at each time-delay step, is shown in figure \ref{W_spec_fig} (b). The clear oscillatory behaviour of the valence band peak in the streaking trace confirms that the photoemitted electron wavepacket is of sub-cycle duration. The electric field at the WO$_{3}$ surface, retrieved from the streaking trace, is shown in figure \ref{W_spec_fig} (d). The accuracy of the retrieved field is confirmed by the excellent agreement with waveforms retrieved from separate gas phase streaking measurements performed in neon \cite{witting_sub_2012}.

The photoelectron wavepacket was retrieved from the background-subtracted streaking trace with no Fourier filter applied, using a standard attosecond retrieval technique, frequency resolved optical gating for complete reconstruction of attosecond bursts \cite{mairesse_frequency_2005} (FROG-CRAB) with a principal components generalised projections algorithm \cite{kane_simultaneous_1997} (PCGPA). The duration of the retrieved photoelectron wavepacket was $359_{-25}^{+42}\,$as. Comparing this figure with the incident XUV pulse duration, the wavepacket broadening in WO$_{3}$ is $111_{-42}^{+57}\,$as. The temporal broadening of $90\,$as expected from the range of photoemission depths is within the experimental error of this value. The measured temporal broadening implies a mean free path of $6.0_{-2.3}^{+3.1}\,\mathrm{\AA}$. These results demonstrate that attosecond resolved measurements can be performed using the streaking technique on amorphous films.

\begin{figure}[htb!]
\centering
\includegraphics[scale=0.55]{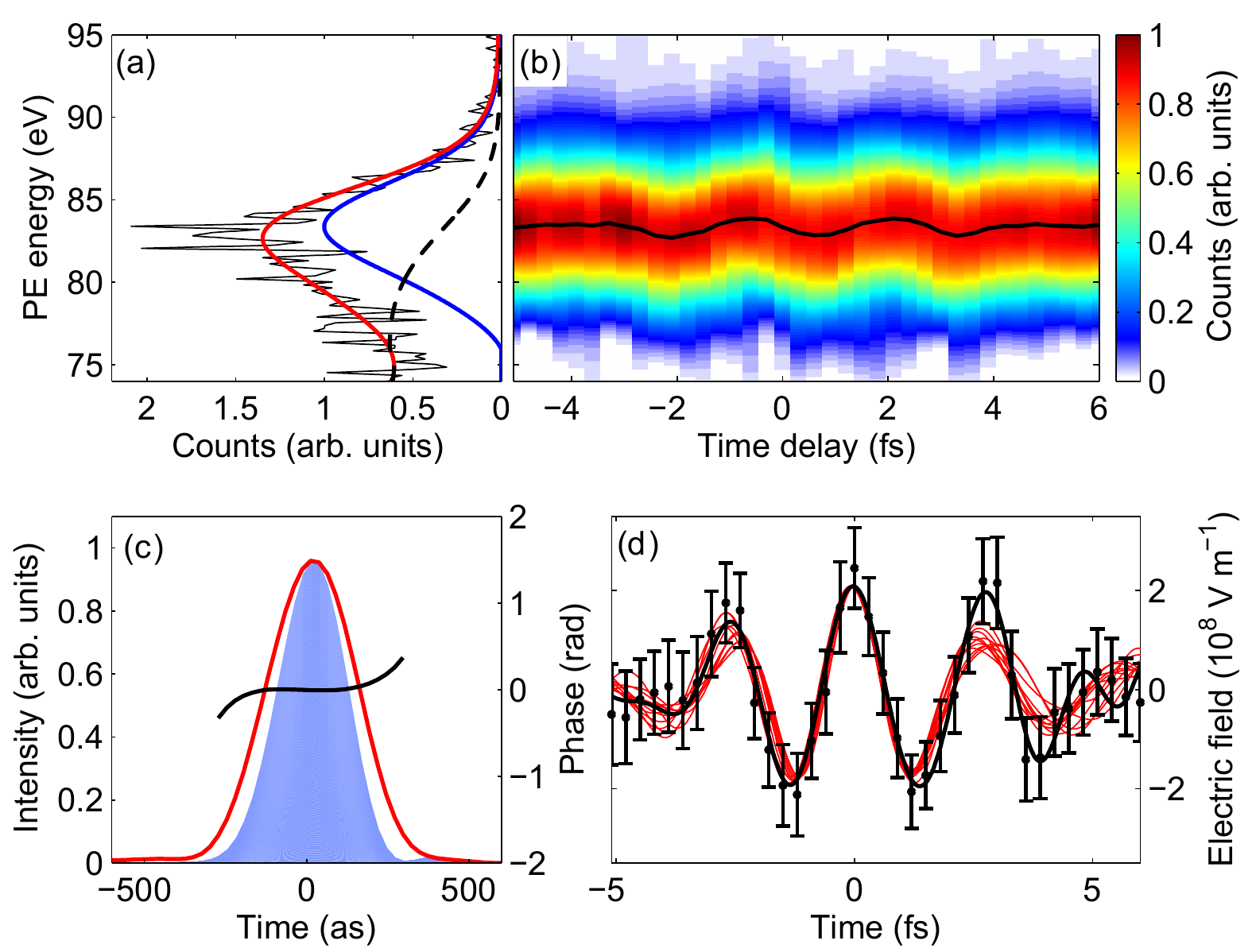}
\caption{\label{Au_spec_fig} (a) Unstreaked valence band PE spectrum of Au. (b) Fourier filtered and background subtracted streaking trace from Au. (c) Photoelectron wavepacket from FROG-CRAB PCGPA algorithm and incident XUV pulse. The retrieved wavepacket has a duration of $319_{-37}^{+43}\,$as. (d) Retrieved electric fields from Au, and from gas phase streaking measurements. The legends are the same as in figure \ref{W_spec_fig}.}
\end{figure}
A streaking measurement on the polycrystalline gold sample is presented in figure \ref{Au_spec_fig}. The streaking trace was acquired using $300\,$as time-delay steps and $40\,$s integration time ($4\times10^{4}$ laser shots) at each step. Again, the photoemission of a sub-cycle electron wavepacket can be confirmed from the oscillatory behaviour of the streaking trace. This demonstration that attosecond streaking can be performed on polycrystalline gold is further evidence that attosecond streaking is applicable to solid samples in general. The retrieved photoelectron wavepacket, shown in figure \ref{Au_spec_fig} (c), has a duration of $319_{-37}^{+43}\,$as. The temporal broadening of $72\,$as expected from the range of photoemission depths agrees well with the experimental value of $71_{-54}^{+58}\,$as. Ths figure corresponds to a $3.9_{-2.9}^{+3.2}\,\mathrm{\AA}$ mean free path. The retrieved near-field at the gold surface, shown in figure \ref{Au_spec_fig} (d), is in agreement with our gas phase streaking measurements.

In conclusion, we have demonstrated that streaking can be used to perform attosecond resolved measurements on solid samples without high degrees of structural order and purity. Attosecond streaking was demonstrated on films of amorphous WO$_{3}$ and polycrystalline Au. The photoemitted electron wavepacket was temporally broadened compared to the incident $248\pm15\,$as XUV pulse by $111_{-42}^{+57}\,$as for WO$_{3}$ and $71_{-54}^{+58}\,$as for Au. These figures are in agreement with the expected range of times taken for electrons emitted at different depths in the solid to travel to the surface. Using the sub-cycle photoemitted wavepackets we were able to make an attosecond-resolved measurement of the electric field at the surface of each sample. Our confirmation that ultrafast optical fields at gold surfaces can be characterised using attosecond streaking constitutes a first step towards the full characterisation of plasmonic near-fields. This should enable ultrafast plasmonic responses in gold nanostructures to be tailored for specific applications in areas such as ultrafast opto-electronics and solar cell technology. Our streaking measurements on WO$_{3}$ open the door to unexplored areas of semiconductor thin-film physics, where charge dynamics can occur on attosecond timescales. Finally, for some experiments in attosecond condensed matter physics it will be possible to relax the sample preparation requirements, which should accelerate progress in this field and significantly broaden its scope.

\subsection*{\normalsize Methods}
\footnotesize
\bf{Laser and streaking setup.} \rm A chirped pulse amplification laser system (Femtolasers GmbH, Femtopower HE CEP) was used to generate $28\,$fs pulses with up to $2.5\,$mJ energy, at a $1\,$kHz repetition rate. Pulses with $1\,$mJ energy were delivered to a differentially pumped hollow core fibre pulse compression system, which was used to produce $0.4\,$mJ, sub-$4\,$fs carrier envelope phase stable few-cycle NIR pulses \cite{okell_carrier_2013}. The few-cycle pulses were used for high-harmonic generation (HHG) in neon within the beamline described in \cite{frank_technology_2012}. The resulting co-propagating high-harmonics and NIR pulses were spatially and spectrally filtered into two beams using a Kapton/Zr filter as shown in figure \ref{setup_fig}. The spatial filtering of the NIR, and additional attenuation using a motorised iris, provided sufficient attenuation to prevent sample damage; the peak intensity at the laser focus ($110\,\mu$m measured spot size) was \mbox{$1.3\times10^{10}\,\mathrm{W}\,\mathrm{cm}^{-2}$} for the WO$_{3}$ measurement, and \mbox{$6\times10^{9}\,\mathrm{W}\,\mathrm{cm}^{-2}$} for the Au measurement (determined from the electric fields retrieved from the streaking traces). At these intensities, above threshold photoemission from the NIR in the valence band region was negligible compared to photoemission from the XUV pulse. The co-propagating NIR and high-harmonics entered a UHV surface science chamber with a base pressure of $3\times10^{-9}\,$mbar. A two-part mirror setup, incorporating an $8\,$eV bandwidth multilayer MoSi mirror with a peak reflectivity at $93\,$eV, selected the cutoff harmonics to produce isolated $248\pm15\,$as XUV pulses (where the pulse duration and error are based on the median and range of pulse durations retrieved from eight separate gas phase streaking measurements). A piezo stage with $10\,$as resolution introduced a time-delay between the XUV and NIR pulses. Photoelectrons emitted from the sample surface were detected using a TOF electron spectrometer with a collection angle of $\pm2.4\,^{\circ}$ and an energy resolution of $\Delta\mathcal{E}/\mathcal{E}\approx0.004$.\\

\bf{Streaking field retrieval.} \rm To retrieve the electric field at the surface, the central energy $\mathcal{E}_\mathrm{COM}$ of the valence band in the Fourier filtered and background subtracted streaking trace was first extracted using a centre-of-mass procedure \cite{cavalieri_attosecond_2007}.  The instantaneous vector potential $\boldsymbol{A}$ of the streaking field at the time of XUV photoemission is then related to the energy shift $\Delta\mathcal{E}=\mathcal{E}_\mathrm{COM}-\mathcal{E}_{\mathrm{init}}$ of the photoelectron spectrum by \cite{itatani_attosecond_2002}
\begin{equation}
\boldsymbol{A} = \sqrt{\frac{m_{e}}{2\mathcal{E}_{\mathrm{init}}}}\frac{\Delta\mathcal{E}}{e\,\mathrm{cos}\,\theta}\,\boldsymbol{\hat{A}},
\end{equation}
where $m_{e}$ and $e$ are the electron mass and charge, respectively, $\mathcal{E}_{\mathrm{init}}$ is the inital photoelectron energy, and $\theta$ is the angle between the initial photoemission velocity and the polarisation vector $\boldsymbol{\hat{A}}$ of the streaking field. The electric field $\boldsymbol{E}$ can be found from $\boldsymbol{A}$ using the relation $\boldsymbol{E}=-\partial\boldsymbol{A}/\partial t$. The electric field was band-pass filtered along the time-delay direction using an order $10$ super-Gaussian filter spanning $460\,$--$1100\,$nm, to remove noise components lying outside the laser spectrum.

\bf{Photoelectron wavepacket retrieval.} \rm The low collection efficiency ($8.8\times10^{-4}$) of our TOF limited the signal-to-noise of the streaking measurements. Significantly higher collection efficiencies (up to $7.6\times10^{-2}$, corresponding to a collection angle of $\pm22.5\,^{\circ}$) are possible with different TOF spectrometer designs (Stefan Kaesdorf Geraete fuer Forschung und Industrie). To allow an estimate of the error on the retrieved wavepacket duration associated with the noise in the streaking traces (which is predominantly from counting statistics), the FROG-CRAB retrieval was performed on the background subtracted streaking trace without any Fourier filtering applied. The Au and WO$_{3}$ traces were interpolated onto $108\times108$ and $134\times134$ grids, respectively, and $2.5\times10^{3}$ iterations of the FROG-CRAB PCGPA\cite{mairesse_frequency_2005,kane_simultaneous_1997} algorithm were performed. The FROG-errors of the retrieved streaking traces were $0.10$ for Au, and $0.05$ for WO$_{3}$. The algorithm errors in the retrieved wavepacket temporal intensity, phase, and duration were calculated from the reconstructed FROG-CRAB traces following the approach in \cite{goulielmakis_single_2008}. The features in the streaking trace with the lowest signal-to-noise are those at the low energy and high energy extrema of the valence band. We estimated the error in the wavepacket duration resulting from these extrema by varying the spectral window size of the input experimental trace into the FROG-CRAB algorithm from $18-22\,$eV (centred around the valence band peak). The error associated with the uncertainty in the secondary electron background was estimated by using two independent methods for its calculation. The overall wavepacket durations and experimental errors were taken as the median and range of retrieved pulse durations for all combinations of spectral window and secondary electron backgrounds.

\subsection*{Acknowledgements}

We gratefully acknowledge financial support from the following sources: EPSRC grants EP/I032517/1, EP/F034601/1, EP/E028063/1; the EPSRC Active Plasmonics Programme; an EPSRC Doctoral Prize Fellowship for \mbox{W. A. O.}; ERC ASTEX project 290467; the Leverhulme Trust; the Landesexzellenzcluster ``Frontiers in Quantum Photon Science''; the Joachim Herz Stiftung; the Excellence Initiative within the project KIT-Nanolab@ANKA. The work was partly carried out with the support of the Karlsruhe Nano Micro Facility, a Helmholtz Research Infrastructure at KIT. We thank \mbox{A. Gregory}, \mbox{P. Ruthven}, and \mbox{S. Parker} for expert technical support, \mbox{J. Overbuschmann} for fabrication of the WO$_{3}$ sample, N. Powell for producing the setup figure, and \mbox{S. Han} and \mbox{B. Buades} for their helpful contributions.

\subsection*{Author contributions}

J. W. G. T., J. P. M., S. A. M. and T. U. concieved the ideas for the experiments. W. A. O., C. A. A. and T. U. designed and built the experimental setup. W. A. O., T. W. and D. F. performed the streaking measurements. W. A. O. and T. W. analyzed the data. D. Y. L., Y. S., M. R. and S. A. M. guided the nanoplasmonics aspects of the work and fabricated the gold sample. J. H., T. U., S. I., A. S., M. B. and S. S. characterized the samples. All authors contributed to the interpretation of the results and preparation of the manuscript.

\subsection*{Additional information}
\bf\normalsize Competing financial interests: \rm The authors declare no competing financial interests.

\end{document}